# Light-Induced Ferromagnetism in Moiré Superlattices

Xi Wang[1,2], Chengxin Xiao[3,4], Heonjoon Park[1], Jiayi Zhu[1], Chong Wang[8], Takashi Taniguchi[5], Kenji Watanabe[6], Jiaqiang Yan[7], Di Xiao[8,1,9], Daniel R. Gamelin[2], Wang Yao[3,4,*], and Xiaodong Xu[1,8,*]

[1]Department of Physics, University of Washington, Seattle, WA, USA
[2]Department of Chemistry, University of Washington, Seattle, WA, USA
[3]Department of Physics, University of Hong Kong, Hong Kong, China
[4]HKU-UCAS Joint Institute of Theoretical and Computational Physics at Hong Kong, China
[5]International Center for Materials Nanoarchitectonics, National Institute for Materials Science, Tsukuba, Ibaraki 305-0044, Japan
[6]Research Center for Functional Materials, National Institute for Materials Science, Tsukuba, Ibaraki 305-0044, Japan
[7]Materials Science and Technology Division, Oak Ridge National Laboratory, Oak Ridge, Tennessee, 37831, USA
[8]Department of Materials Science and Engineering, University of Washington, Seattle, WA, USA
[9]Pacific Northwest National Laboratory, Richland, Washington, United States

Correspondence to: wangyao@hku.hk; xuxd@uw.edu

**Abstract:** Many-body interactions between carriers lie at the heart of correlated physics. The ability to tune such interactions would open the possibility to access and control complex electronic phase diagrams on demand. Recently, moiré superlattices formed by two-dimensional materials have emerged as a promising platform for quantum engineering such phenomena[1-3]. The power of the moiré system lies in the high tunability of its physical parameters by tweaking layer twist angle[1-3], electrical field[4-6], moiré carrier filling[7-11], and interlayer coupling[12]. Here, we report that optical excitation can drastically tune the spin-spin interactions between moiré trapped carriers, resulting in ferromagnetic order in $WS_2/WSe_2$ moiré superlattices over a small range of doping at elevated temperatures. Near the filling factor $v = -1/3$ (i.e., one hole per three moiré unit cells), as the excitation power at the exciton resonance increases, a well-developed hysteresis loop emerges in the reflective magnetic circular dichroism (RMCD) signal as a function of magnetic field, a hallmark of ferromagnetism. The hysteresis loop persists down to charge neutrality, and its shape evolves as the moiré superlattice is gradually filled, indicating changes of magnetic ground state properties. The observed phenomenon points to a mechanism in which itinerant photo-excited excitons mediate exchange coupling between moiré trapped holes. This exciton-mediated interaction can be of longer range than direct coupling between moiré trapped holes[9], and thus magnetic order can arise even in the dilute hole regime under optical excitation. This discovery adds a new and dynamic tuning knob to the rich many-body Hamiltonian of moiré quantum matter[13-19].

**Main text**

Moiré superlattices formed by semiconducting transition metal dichalcogenides (TMDs) represent an emerging platform to explore correlated effects with high tunability[7-10,20-23]. Combined with strong Coulomb interactions, triangular moiré geometry, strong spin-orbit coupling, and isolated flat electronic bands, the TMD heterobilayer is an ideal playground to test tunable many-body Hamiltonians. Indeed, correlated insulating states at both integer and fractional moiré miniband fillings have been demonstrated experimentally[7-11]. Local magnetic moments were also observed near one hole per moiré until cell filling (i.e., $v = -1$) that couple antiferromagnetically[9], attributed to the well-known solution of the Hubbard model with a triangular lattice at half filling. Theoretically, the TMD moiré platform offers an opportunity to investigate classic toy models with triangular or hexagonal geometry for exploring strongly correlated physics.[13-19] By varying the onsite Coulomb interaction $U$ and the nearest neighbor hopping parameter $t$, many-body phase diagrams with various insulating, metallic, and exotic magnetic and topological states have been predicted[13-19].

To experimentally realize these states, the ability to tailor the interactions between moiré trapped spins *in situ*, and thus the many-body Hamiltonian is highly desirable. One possible approach is to utilize optical excitation. In fact, optical excitation of strongly correlated quantum materials can have profound effects on the many-body electronic states, such as light induced superconductivity[24], hidden charge density wave order[25], etc. In the context of TMD heterobilayers, the exciton stands out as a promising candidate, which is itself of great interest for the strong light-matter interactions, unique spin-valley coupled optical effects[26], and moiré superlattice physics[27-31]. These optical excitations are already exploited as effective probes of correlated insulating and generalized Wigner crystal states in moiré superlattices[7-9,20-23]. However, the impact of optical excitation on many-body electronic states, i.e., non-equilibrium effects, in the strongly correlated moiré superlattice, remains unexplored.

Here, we report the discovery that optical excitation can drastically enhance spin-spin exchange interaction $J$ between moiré trapped carriers and thus triggers the emergence of ferromagnetic order in the $WS_2/WSe_2$ moiré superlattice. This finding points to a profound and dynamic control of the moiré many-body Hamiltonian. Dual gated $WS_2/WSe_2$ samples were fabricated to allow independent control of carrier doping and vertical electrical field (Figure 1a, see methods). Figure 1b shows a piezoresponse force microcopy image of a representative device. A homogeneous triangular moiré superlattice is observed with a moiré wavelength of about 7.5 nm, consistent with near zero-degree twist angle alignment. Optical reflection measurements as a function of doping show typical moiré excitons with moiré miniband filling effects (Fig. 1c), as reported previously[7,9,29]. The filling factor $v$ is labeled on the right axis (see Extended Data Fig. 1 for the filling factor assignment).

To probe the magnetic properties, we performed reflective magnetic circular dichroism (RMCD) measurements. Figure 1d plots the RMCD intensity as a function of $v$ and magnetic field $\mu_oH$ applied in the out of plane direction. We mainly focus on $-\frac{3}{2} \leq v \leq \frac{2}{3}$, where the magnetic response is most appreciable. The data are taken with the laser excitation in resonance with the $WSe_2$ A exciton (see wavelength dependent data in Extended Data Fig. 2). The excitation power is 200 nW and the experimental temperature is 1.6 K (see Methods). Fig. 1d displays a rich $v$ dependent magneto-optical response. These effects are illustrated by the RMCD line traces at selected filling factors in Fig. 1e. At $v = -1$, i.e., one hole per moiré unit cell, RMCD shows a superparamagnetic-like response and saturates for $|\mu_o H|$ above 1 T. This demonstrates the existence of local magnetic moments. The superparamagnetic-like response is robust for $v$ near -1 (see Extended Data Fig. 3 for selected RMCD line traces). The observation is consistent with previous report of intralayer exciton Zeeman splittings[9].

Remarkably, when hole doping becomes significantly smaller, a magnetic hysteresis loop starts to appear in the RMCD signal vs $\mu_o H$, a signature of ferromagnetic order (See Extended Data Fig. 4 for filling factor dependent loop width). The RMCD lineshape is very sensitive to doping. In a narrow doping regime near $v = -1/3$, the RMCD signal vs $\mu_o H$ shows typical ferromagnetic behavior, distinct from linear valley Zeeman effect. It possesses a hysteresis loop with spin flip transitions around $\mu_o H_c = \pm 11$ mT, above which the signal saturates (Fig. 1e and Extended Data Fig. 5). When the doping density further reduces, the RMCD line traces evolves into a heartbeat lineshape, and eventually vanishes as electron doping dominates. The emergent sharp RMCD features near zero magnetic field, which are associated with ferromagnetic states at low hole doping regime, are the focus of this paper.

We first present the results near $v = -1/3$. Figure 2a shows the RMCD signal as a function of excitation power $P$. When $P$ is smaller than 16 nW, the RMCD signal vs $\mu_o H$ vanishes (after subtracting a constant background) and behaves as a featureless straight line. A hysteresis loop emerges as $P$ is increased above a critical threshold. The amplitude of the remnant RMCD signal at zero field grows as $P$ increases, and eventually saturates (Fig. 2b). In comparison, the RMCD response and its saturation value at $v = -1$ have little excitation power dependence (Extended Data Fig. 3). This is consistent with the attribution of the RMCD response at $v = -1$ to the intrinsic magnetic interactions[9]. At low filling factors, however, the intrinsic magnetic interaction is significantly weaker because of larger hole-hole distances compared to the $v = -1$ case. Thus, the peculiar power dependent RMCD responses at the fractional filling factor $v = -1/3$ signify the emergence of ferromagnetic order by optically induced long-range spin-spin interaction.

The hysteresis loop width exhibits negligible dependence on the optical excitation power. This can be understood that at temperatures much lower than the Curie temperature, the loop width is mainly determined by the magnetic anisotropy. Our measurements thus suggest that magnetic anisotropy is intrinsic to this moiré system. We note that the width of the hysteresis loop depends on the magnetic field sweeping rate (Extended data Fig. 5). Slower sweeping leads to a narrower hysteresis loop width. This result reflects slow magnetic domain dynamics in the magnetization flipping process. Here, all magnetic field sweep results are taken with a sweep rate of 14 Gauss/second, below which the hysteresis loop width is nearly independent of the sweep rate.

The ferromagnetic order is further confirmed by temperature dependent RMCD measurements, shown in Fig. 2c, measured at $P = 103$ nW. The hysteresis loop width reduces as the temperature increases, as shown in Fig. 2d. There is a 5 mT instrument offset above 10 K until the noise level exceeds the signal above ~45K. Similar temperature dependence is also observed at other filling factors (Extended Data Fig. 4). The effective Curie temperature is determined to be about 8 K. In the mean field limit, this gives rise to an estimated exchange interaction $J$ on the order of -1 meV (with the Hamiltonian $H$ in the form of $J S_1 \cdot S_2$, see methods). Note that the main experimental results are reproduced in an additional device (Extended Data Fig. 6).

At $v = -1/3$, photoluminescence measurements show a correlated insulating state (Extended Data Fig. 1). This is consistent with previous reports that a generalized Wigner crystal forms with holes arranged in a triangular lattice[7-10,20-23]. Monte Carlo simulations also suggest that this $v = -1/3$ state is relatively robust compared to other fractional filling charge orders (Extended Data Fig. 1). The period of the charge order at $v = -1/3$ (~13 nm) is $\sqrt{3}$ times larger than that at $v = -1$. Thus, the intrinsic spin-spin interaction of the former should be much smaller compared to the latter. This is manifested by the vanishing RMCD signal taken at very low excitation power (Fig. 2a), which implies negligible intrinsic magnetic interaction at such a low filling. The power dependent RMCD signal highlights the drastic optical tuning of the spin-spin interaction strength $J$, and hence of the magnetic phase in the moiré superlattice (Fig. 2d, inset).

When the hole doping density reduces and $v$ deviates from -1/3, the line trace of RMCD signal versus $\mu_o H$ quickly evolves into heartbeat-like shapes with a narrow width (<100 mT), as shown in Fig. 1e. Figure 3a plots the zoom-in RMCD intensity in the low doping regime. The signal reaches maximum near $v = -1/7$, and is still appreciable at nominal charge neutrality. Although the total net charge of the system vanishes at the charge neutrality point, inevitable defects as well as optically created photo-carriers can give rise to puddles of dilute holes trapped in the moiré potential of the $WSe_2$ layer. As electron doping increases, the RMCD signal reduces and eventually vanishes. The signal peaking at $v = -1/7$ is likely due to the formation of a correlated insulating state, formed by holes arranged in a triangular lattice (Extended Data Fig. 1).

Figure 3b shows the RMCD signal as a function of excitation power $P$ at $v = -1/7$. It starts with a slightly tilted straight line without any hysteresis, which is the typical magnetic-optical response of TMDs without magnetic order. As $P$ increases, the heartbeat lineshape emerges with a hysteresis loop as the magnetic field is swept back and forth. The RMCD amplitude, defined as the peak-to-valley value of the heartbeat lineshape, increases with $P$ and eventually saturates (Fig. 3c). These results demonstrate optically induced ferromagnetism in the dilute hole doping regime.

The above experimental observations establish optical creation and tuning of the magnetic order, otherwise absent in dark conditions, in the $WS_2/WSe_2$ moiré superlattices. We note that the RMCD probe of the magnetic response is through resonant excitation of the $WSe_2$ exciton, upon which ultrafast interlayer charge transfer leads to formation of long-lived interlayer excitons[32,33]. Experiments on $MoSe_2/WSe_2$ moiré superlattices have shown that at the low excitation limit (< 10 nW), these excitons are mostly trapped by the moiré potentials, giving rise to narrow photoluminescence lines. With increasing excitation power, the steady state population of itinerant (or untrapped) interlayer excitons grows significantly, manifested by a broad luminescence background[27]. Our observations therefore point to the role of these optically injected itinerant excitons in aiding the magnetic interactions between holes trapped in moiré potentials. Indeed, spin-spin interactions between two spatially separated holes can arise as a second-order process mediated by their exchange interaction with a common itinerant exciton, as schematically illustrated in Fig. 3d. Similar role of exciton has been theoretically explored in a virtual excitation regime, where a fourth-order perturbation process leads to short-ranged exchange interaction of localized spins in III-V semiconductors[34], which may also result in magnetic order[35].

We found that this exciton mediated hole-hole exchange interaction is ferromagnetic (see Methods). The strength of the exciton mediated exchange interaction depends on the separation of moiré trapped holes, exciton density, and effective temperature of the exciton gas which can in general be different from the lattice temperature. Figure 3e plots the calculated $J$ as a function of hole separation and exciton density, at an exciton temperature of 20 K (see Extended Data Fig. 7 for other exciton temperatures). We found that $J$ remains significant at a hole-hole distance of 20 nm. At a reasonable exciton density of $10^{11}$ cm$^{-2}$ (see Methods), its strength can reach the order of 0.5 meV (Fig. 3e). We note that a quantitative comparison between the calculation and experiment is challenging for non-equilibrium many-body problems. For instance, the RMCD response does not resemble the typical magnetization curve even qualitatively, subject to the influence of the excitonic lineshape. The exact origin of the lineshapes, for example the heartbeat at dilute hole doping regime, is not clear. It might reflect the change of the magnetic ground state properties upon doping. Nevertheless, the observed RMCD hysteresis upon the forward and backward sweeping of magnetic field is unambiguous evidence of magnetic order, which is consistent with our calculation that excitons can effectively establish ferromagnetic exchange interactions in dilute moiré hole gas.

We further performed temperature dependent RMCD measurements at $v = -1/7$ as a function of optical excitation power, ranging from 26 to 253 nW. Figure 4a illustrates the results at select excitation powers. We define a critical temperature $T_C$, above which the heartbeat amplitude vanishes (below the noise level 0.2%), as for the magnetic response. Using 253 nW optical excitation as an example, the heartbeat lineshape remains strong up to about 40 K. Fig. 4a displays a strong dependence of $T_C$ on the optical excitation power. To further highlight this effect, we plot the extracted amplitude of the RMCD signal as a function of excitation power and temperature in Fig. 4b (see Extended Data Fig. 8 for line cuts). These data show that once the optical excitation power is large enough to introduce magnetic order, $T_C$ can be optically tuned from about 20 K to 45 K.

$T_C$ also exhibits strong dependence on the filling factors. Fig. 4c shows RMCD signal versus magnetic field as a function of temperature at select $v$ (Extended Data Fig. 9). We then plot the extracted RMCD amplitude in Fig. 4d. This figure highlights that for the magnetic states within $-2/3 < v < 1/3$, $T_C$ peaks near correlated insulating states at fractional fillings and drops sharply once $v$ deviates from these states (e.g., -1/2 and -1/7, and see -1/3 in Extended Data Fig. 10). While the correlated insulating states are not required for the formation of light induced magnetic states, these observations suggest that the latter is further stabilized with the formation of charge order, where the spin fluctuations among moiré spins are minimized. The optical excitation power and filling factor dependent results underscore the broad tunability of the moiré many-body Hamiltonian, which may lead to the creation of exotic states unique to moiré superlattices. We envision several immediate exciting possibilities, such as exploring the effect of optical orientation of exciton spins on the formation of magnetic states[36,37], electric field tuning effects, transient magnetism, optically controlled topological phase transitions, magneto-exciton polaritons in the absence of magnetic field, and many-spin entangled states for optically driven quantum information[34].

**Methods:**

**Sample fabrication.**
Mechanically exfoliated monolayers of WS$_2$ and WSe$_2$ are stacked using a dry-transfer technique on hexagonal boron nitride (BN) and graphene back gate, with prefabricated Platinum contacts. The crystal orientation of the individual monolayers is first determined by linear-polarization resolved second-harmonic generation before transfer. The alignment angle of $WS_2$ and $WSe_2$ is double checked by piezoresponse force microscopy (PFM) during transfer before encapsulating with top BN and graphite. The BN encapsulation (20 – 40 nm) provides an atomically smooth substrate. PFM is performed on a Bruker Dimension Icon AFM. Platinum-Iridium coated, electrically conductive AFM probes (Bruker's SCM-PIT-V2) with a force constant of ~3 N m$^{-1}$ are used. An AC bias is applied between the tip and sample that induces a periodic deformation of the sample whose amplitude and phase give local information on the electromechanical response. Small AC bias magnitudes (<300 mV) with resonance frequencies in the range of ~ 700 kHz are applied for lateral PFM.

**Optical measurements.**
All PL and differential optical reflection measurements are performed in a home-built confocal optical microscope in the reflection geometry. The sample is mounted in an exchange-gas cooled cryostat (attoDRY 2100) equipped with a 9 T superconducting magnet in Faraday configuration. The sample temperature is kept at 1.6 K unless otherwise specified. SuperK Extreme Supercontinuum White Light laser (pulse duration around 2 ns; repetition rate around 78 MHz) and LLTF tunable high contrast filter with selected wavelength 739.2 nm are used to excite the sample at normal incidence, unless otherwise specified. The power is specified in the manuscript and corresponding figure captions. The Gaussian profile of the excitation beam has FWHM around 700 nm. We also used the same laser with white light emission for $\Delta R/R$ with 600-nm long pass and 800-nm short pass (average power 15 nW). Reflectance and PL signals are dispersed by a diffraction grating (600 grooves per mm) and detected on a silicon CCD camera. The photoluminescence is spectrally filtered from the laser using a long-pass filter before being directed into a spectrometer.

The reflective magnetic circular dichroism (RMCD) measurement was performed with an AC lock-in measurement technique. In brief, an intensity modulated laser beam (at 800 Hz) was linearly polarized 45° to the photoelastic modulator (PEM) slow axis. Transmitting through the PEM, the light was sinusoidally phase modulated at 50.1 kHz, with a maximum retardance of $\lambda/4$. This produces an alternating left and right circularly polarized light at the phase modulation frequency. The modulated light was focused down onto the sample at normal incidence. The optical reflection was then separated from the incidence path via a laser line nonpolarizing beam splitter and projected onto a photodiode, whose signal is sent into two lock-in amplifiers: one is tuned at 50.1 kHz to detect the RMCD, and the other is tuned at the chopper frequency, 800 Hz, to normalize the RMCD signal to laser intensity, reflected off the sample.

**Estimation of interlayer exciton (IX) density.**
We estimate the upper bound of the exciton density under 100 nW pulse excitation (78 MHz, pulse duration 2 ns) with photon energy in resonance with $WSe_2$ A exciton (1.677 eV). The optical absorption is mainly from the $WSe_2$ layer which is about 10%.[38] Approximately 500 photons, and thus 500 excitons, are excited per pulse. We normalize the exciton population over the Gaussian profile of the excitation beam (FWHM 700 nm). Assuming near unity intralayer to interlayer exciton conversion efficiency, and due to the short duration of the pulse compared to the exciton lifetime, the IX density is about $10^{11}$ cm$^{-2}$ immediately after a pulse.

**Estimation of filling factor based on doping density.**

The doping densities in the heterobilayer are determined from the applied gate voltages based on the parallel-plate capacitor model. The thickness of BN is determined by atomic force microscopy. Both top and bottom BN flakes of the device presented in the main text are 37 nm (top BN) and 38 nm (bottom BN). The doping density is calculated as $C_t\Delta V_t + C_b\Delta V_b$, where $C_t$ and $C_b$ are the capacitance of the top and bottom gates. $\Delta V_t$ and $\Delta V_b$ are the applied gate voltages relative to valence/conduction band edges. The geometric capacitance is calculated with dielectric constant $\varepsilon_{hBN} \sim 3$. The moiré lattice constant is obtained directly from PFM. The filling factor is estimated based on the doping density and size of the moiré unit cell, which is then compared with the assignment of integer filling factors based on differential reflectance and PL.

**Calculation of exciton medicated magnetic exchange interaction.**

**Coulomb exchange between exciton and trapped hole.** The Coulomb interactions between electrons, between holes, and between electron and hole are expressed as[39]

$$\begin{aligned}
H_{ee} &= \tfrac{1}{2}\sum_{i,j,k,l}\langle ij|V|kl\rangle e_i^\dagger e_j^\dagger e_l e_k \\
H_{hh} &= \tfrac{1}{2}\sum_{\alpha,\beta,\gamma,\delta}\langle \alpha\beta|V|\gamma\delta\rangle h_\delta^\dagger h_\gamma^\dagger h_\alpha h_\beta \\
H_{eh} &= \sum_{i,j,\alpha,\beta}[\langle i\beta|V|\alpha j\rangle - \langle i\beta|V|j\alpha\rangle]\, e_i^\dagger h_\alpha^\dagger h_\beta e_j
\end{aligned} \quad (1)$$

Here, $e_i^\dagger$ and $h_\alpha^\dagger = e_\alpha$ denote electron and hole creation operators, respectively. Their matrix elements between three-particle states consisting of one electron and two holes are

$$\begin{aligned}
&\langle c_{\mathbf{e}'} v_{\mathbf{h}_1'} v_{\mathbf{h}_2'}|H_{hh}|c_\mathbf{e} v_{\mathbf{h}_1} v_{\mathbf{h}_2}\rangle \\
&= \tfrac{1}{2}\sum_{\alpha,\beta,\gamma,\delta}\langle \alpha\beta|V|\gamma\delta\rangle\langle 0|h_{\mathbf{h}_2'} h_{\mathbf{h}_1'} e_{\mathbf{e}'} h_\delta^\dagger h_\gamma^\dagger h_\alpha h_\beta e_\mathbf{e}^\dagger h_{\mathbf{h}_1}^\dagger h_{\mathbf{h}_2}^\dagger|0\rangle \\
&= \tfrac{1}{2}\sum_{\alpha,\beta,\gamma,\delta}\langle \alpha\beta|V|\gamma\delta\rangle\delta_{\mathbf{e}'\mathbf{e}}\langle 0|h_{\mathbf{h}_2'}(\delta_{\mathbf{h}_1'\delta} - h_\delta^\dagger h_{\mathbf{h}_1'})h_\gamma^\dagger h_\alpha(\delta_{\mathbf{h}_1\beta} - h_{\mathbf{h}_1}^\dagger h_\beta)h_{\mathbf{h}_2}^\dagger|0\rangle \\
&= \tfrac{1}{2}\delta_{\mathbf{e}'\mathbf{e}}\big[\langle v_{\mathbf{h}_1} v_{\mathbf{h}_2}|V|v_{\mathbf{h}_1'} v_{\mathbf{h}_2'}\rangle + \langle v_{\mathbf{h}_2} v_{\mathbf{h}_1}|V|v_{\mathbf{h}_2'} v_{\mathbf{h}_1'}\rangle - \langle v_{\mathbf{h}_1} v_{\mathbf{h}_2}|V|v_{\mathbf{h}_2'} v_{\mathbf{h}_1'}\rangle - \langle v_{\mathbf{h}_2} v_{\mathbf{h}_1}|V|v_{\mathbf{h}_1'} v_{\mathbf{h}_2'}\rangle\big] \\
&= \delta_{\mathbf{e}'\mathbf{e}}\big(\langle v_{\mathbf{h}_1} v_{\mathbf{h}_2}|V|v_{\mathbf{h}_1'} v_{\mathbf{h}_2'}\rangle - \langle v_{\mathbf{h}_1} v_{\mathbf{h}_2}|V|v_{\mathbf{h}_2'} v_{\mathbf{h}_1'}\rangle\big)
\end{aligned} \quad (2)$$

$$\begin{aligned}
&\langle c_{\mathbf{e}'} v_{\mathbf{h}_1'} v_{\mathbf{h}_2'}|H_{eh}|c_\mathbf{e} v_{\mathbf{h}_1} v_{\mathbf{h}_2}\rangle \\
&= \sum_{\alpha,\beta,i,j}[\langle i\beta|V|\alpha j\rangle - \langle i\beta|V|j\alpha\rangle]\,\langle 0|h_{\mathbf{h}_2'} h_{\mathbf{h}_1'} e_{\mathbf{e}'} e_i^\dagger h_\alpha^\dagger h_\beta e_j e_\mathbf{e}^\dagger h_{\mathbf{h}_1}^\dagger h_{\mathbf{h}_2}^\dagger|0\rangle \\
&= \sum_{\alpha,\beta,i,j}[\langle i\beta|V|\alpha j\rangle - \langle i\beta|V|j\alpha\rangle]\,\delta_{\mathbf{e}'i}\delta_{\mathbf{e}j}\langle 0|h_{\mathbf{h}_2'}(\delta_{\mathbf{h}_1'\alpha} - h_\alpha^\dagger h_{\mathbf{h}_1'})(\delta_{\mathbf{h}_1\beta} - h_{\mathbf{h}_1}^\dagger h_\beta)h_{\mathbf{h}_2}^\dagger|0\rangle \\
&= \big[\langle c_{\mathbf{e}'} v_{\mathbf{h}_1}|V|v_{\mathbf{h}_1'} c_\mathbf{e}\rangle - \langle c_{\mathbf{e}'} v_{\mathbf{h}_1}|V|c_\mathbf{e} v_{\mathbf{h}_1'}\rangle\big]\delta_{\mathbf{h}_2'\mathbf{h}_2} + \big[\langle c_{\mathbf{e}'} v_{\mathbf{h}_2}|V|v_{\mathbf{h}_2'} c_\mathbf{e}\rangle - \langle c_{\mathbf{e}'} v_{\mathbf{h}_2}|V|c_\mathbf{e} v_{\mathbf{h}_2'}\rangle\big]\delta_{\mathbf{h}_1'\mathbf{h}_1} \\
&\quad - \big[\langle c_{\mathbf{e}'} v_{\mathbf{h}_2}|V|v_{\mathbf{h}_1'} c_\mathbf{e}\rangle - \langle c_{\mathbf{e}'} v_{\mathbf{h}_2}|V|c_\mathbf{e} v_{\mathbf{h}_1'}\rangle\big]\delta_{\mathbf{h}_2'\mathbf{h}_1} - \big[\langle c_{\mathbf{e}'} v_{\mathbf{h}_1}|V|v_{\mathbf{h}_2'} c_\mathbf{e}\rangle - \langle c_{\mathbf{e}'} v_{\mathbf{h}_1}|V|c_\mathbf{e} v_{\mathbf{h}_2'}\rangle\big]\delta_{\mathbf{h}_1'\mathbf{h}_2}
\end{aligned} \quad (3)$$

Here the bold character denotes all the degrees of freedom including momentum $\mathbf{k}$ and spin-valley index $\sigma$, e.g. $\mathbf{e} \equiv (\mathbf{k}_e, \sigma_e)$). $\langle c_{\mathbf{k}_1,\sigma_1} v_{\mathbf{k}_2,\sigma_2}|V|c_{\mathbf{k}_3,\sigma_3} v_{\mathbf{k}_4,\sigma_4}\rangle$ and $\langle v_{\mathbf{k}_1,\sigma_1} v_{\mathbf{k}_2,\sigma_2}|V|v_{\mathbf{k}_3,\sigma_3} v_{\mathbf{k}_4,\sigma_4}\rangle$ are the two-body matrix elements in momentum space with the following expression[40]

$$\begin{aligned}
&\langle c_{\mathbf{k}_1,\sigma_1} v_{\mathbf{k}_2,\sigma_2}|V|c_{\mathbf{k}_3,\sigma_3} v_{\mathbf{k}_4,\sigma_4}\rangle = \int d\mathbf{r}_1 d\mathbf{r}_2\, \psi^*_{\mathbf{k}_1,\sigma_1,c}(\mathbf{r}_1)\psi^*_{\mathbf{k}_2,\sigma_2,v}(\mathbf{r}_2)V(\mathbf{r}_1-\mathbf{r}_2)\psi_{\mathbf{k}_3,\sigma_3,c}(\mathbf{r}_1)\psi^*_{\mathbf{k}_4,\sigma_4,v}(\mathbf{r}_2) \\
&= \delta_{\mathbf{k}_1-\mathbf{k}_3,\mathbf{k}_2-\mathbf{k}_4}\sum_\mathbf{G}\frac{V(\mathbf{k}_1-\mathbf{k}_3+\mathbf{G})}{A}\langle u_{\mathbf{k}_1,\sigma_1,c}|e^{i\mathbf{G}\cdot\mathbf{r}}|u_{\mathbf{k}_3,\sigma_3,c}\rangle\langle u_{\mathbf{k}_2,\sigma_2,v}|e^{-i\mathbf{G}\cdot\mathbf{r}}|u_{\mathbf{k}_4,\sigma_4,v}\rangle \\
&\langle v_{\mathbf{k}_1,\sigma_1} v_{\mathbf{k}_2,\sigma_2}|V|v_{\mathbf{k}_3,\sigma_3} v_{\mathbf{k}_4,\sigma_4}\rangle = \int d\mathbf{r}_1 d\mathbf{r}_2\, \psi^*_{\mathbf{k}_1,\sigma_1,v}(\mathbf{r}_1)\psi^*_{\mathbf{k}_2,\sigma_2,v}(\mathbf{r}_2)V(\mathbf{r}_1-\mathbf{r}_2)\psi^*_{\mathbf{k}_3,\sigma_3,v}(\mathbf{r}_1)\psi^*_{\mathbf{k}_4,\sigma_4,v}(\mathbf{r}_2) \\
&= \delta_{\mathbf{k}_1+\mathbf{k}_2,\mathbf{k}_3+\mathbf{k}_4}\sum_\mathbf{G}\frac{V(\mathbf{k}_1-\mathbf{k}_3+\mathbf{G})}{A}\langle u_{\mathbf{k}_3,\sigma_3,v}|e^{i\mathbf{G}\cdot\mathbf{r}}|u_{\mathbf{k}_1,\sigma_1,v}\rangle\langle u_{\mathbf{k}_4,\sigma_4,v}|e^{-i\mathbf{G}\cdot\mathbf{r}}|u_{\mathbf{k}_2,\sigma_2,v}\rangle
\end{aligned} \quad (4)$$

where $|u\rangle$ is the periodical part of the Bloch function, $V(\mathbf{k}) = \int d\mathbf{r} e^{i\mathbf{k}\cdot\mathbf{r}} V(\mathbf{r})$ is the Fourier transformation of 2D Coulomb potential $V(\mathbf{r})$ and $A$ is the area of the 2D plane (for box normalization).

The Coulomb exchange interaction between the exciton and the trapped hole is evaluated, in the basis $|\Psi_{\mathbf{k},j\sigma_e\sigma_h\sigma}\rangle = d^\dagger_{j\sigma} a^\dagger_{\mathbf{k}\sigma_e\sigma_h}|0\rangle$, where $|\Psi_{X,\mathbf{k}\sigma_e\sigma_h}\rangle = a^\dagger_{\mathbf{k}\sigma_e\sigma_h}|0\rangle$ is the momentum eigenstate of the interlayer exciton with electron (hole) spin/valley index $\sigma_e$ ($\sigma_h$), and $|\Psi_{h,j\sigma}\rangle = d^\dagger_{j\sigma}|0\rangle$ denotes a trapped hole wave packet centered at $\mathbf{R}_j$. We have $|\Psi_{X,\mathbf{k}\sigma_e\sigma_h}\rangle = \sum_{\mathbf{q}_X} \psi_X(\mathbf{q}_X) e^\dagger_{\mathbf{q}_X+\mathbf{k}/2,\sigma_e} h^\dagger_{\mathbf{q}_X-\mathbf{k}/2,\sigma_h}|0\rangle$ and $|\Psi_{h,j\sigma}\rangle = \sum_{\mathbf{k}_h} e^{i\mathbf{k}_h\cdot\mathbf{R}_j} W_h(\mathbf{k}_h) h^\dagger_{\mathbf{k}_h,\sigma}|0\rangle$ with $W_h(\mathbf{r}_h) = e^{-r_h^2/a_h^2}$ and $\psi_X(\mathbf{r}_{eh}) = e^{-r_{eh}/a_b}$ for the 1s exciton. The corresponding Fourier transformations are $W_h(\mathbf{k}_h) = \frac{a_h^2}{2} e^{-\frac{1}{4}a_h^2 k_h^2}$ and $\psi_X(\mathbf{q}_X) = \frac{a_b^2}{(1+q_X^2 a_b^2)^{3/2}}$. The normalized wave function of the basis state is:

$$|\Psi_{\mathbf{k},j\sigma_e\sigma_h\sigma}\rangle = C \sum_{\mathbf{q}_X,\mathbf{k}_h} \psi_X(\mathbf{q}_X) e^{i\mathbf{k}_h\cdot\mathbf{R}_j} W_h(\mathbf{k}_h) e^\dagger_{\mathbf{q}_X+\mathbf{k}/2,\sigma_e} h^\dagger_{\mathbf{q}_X-\mathbf{k}/2,\sigma_h} h^\dagger_{\mathbf{k}_h,\sigma}|0\rangle \tag{5}$$

with the normalization condition:

$$\langle\Psi_{\mathbf{k},j\sigma_e\sigma_h\sigma}|\Psi_{\mathbf{k},j\sigma_e\sigma_h\sigma}\rangle = \frac{C^2 A^2}{(2\pi)^4} \int d\mathbf{q}_X d\mathbf{k}_h |W_h(\mathbf{k}_h)|^2 |\psi_X(\mathbf{q}_X)|^2 = 1 \tag{6}$$

For the interlayer exciton and trapped hole in the type-II heterobilayer, holes are layer separated from the electron component of the exciton. Therefore, we can neglect the terms that relies on wavefunction overlap between the electron and hole (second, fourth, sixth and eighth term in Eq. (3)). The Coulomb exchange interaction between the exciton and the trapped hole are then contributed by the second term in Eq. (2) from the hole-hole interaction, and the fifth and seventh terms in Eq. (3) from the electron-hole interaction,

$$\begin{aligned}
&\langle\Psi_{\mathbf{k}',j\sigma'_e\sigma'_h\sigma'}|H_{hh}|\Psi_{\mathbf{k},j\sigma_e\sigma_h\sigma}\rangle_{second} \\
&= -\delta_{\sigma'_e\sigma_e} C^2 \sum_{\mathbf{q}_X \mathbf{k}_h \mathbf{q}'_X \mathbf{k}'_h} e^{i(\mathbf{k}_h - \mathbf{k}'_h)\cdot\mathbf{R}_j} W_h^*(\mathbf{k}'_h) \psi_X^*(\mathbf{q}'_X) W_h(\mathbf{k}_h) \psi_X(\mathbf{q}_X) \delta_{\mathbf{q}'_X + \mathbf{k}'/2, \mathbf{q}_X + \mathbf{k}/2} \\
&\quad \delta_{\mathbf{k}' - \mathbf{k}'_h, \mathbf{k} - \mathbf{k}_h} \sum_{\mathbf{G}} \frac{V(\mathbf{q}_X - \mathbf{k}/2 - \mathbf{k}'_h + \mathbf{G})}{A} \langle u_{\mathbf{k}'_h,v}|e^{i\mathbf{G}\cdot\mathbf{r}}|u_{\mathbf{q}_X-\mathbf{k}/2,v}\rangle \langle u_{\mathbf{q}'_X-\mathbf{k}'/2,v}|e^{-i\mathbf{G}\cdot\mathbf{r}}|u_{\mathbf{k}_h,v}\rangle \\
&\approx -\delta_{\sigma'_e\sigma_e} \delta_{\sigma'_h\sigma} \delta_{\sigma_h\sigma'} e^{-i(\mathbf{k}-\mathbf{k}')\cdot\mathbf{R}} C^2/A \\
&\quad \sum_{\mathbf{q}_X \mathbf{k}_h} W_h^*(\mathbf{k}_h - \mathbf{k} + \mathbf{k}') \psi_X^*\left(\mathbf{q}_X + \frac{\mathbf{k}-\mathbf{k}'}{2}\right) W_h(\mathbf{k}_h) \psi_X(\mathbf{q}_X) V\left(\mathbf{q}_X + \frac{\mathbf{k}}{2} - \mathbf{k}' - \mathbf{k}_h\right)
\end{aligned} \tag{7}$$

$$\begin{aligned}
&\langle\Psi_{\mathbf{k}',j\sigma'_e\sigma'_h\sigma'}|H_{eh}|\Psi_{\mathbf{k},j\sigma_e\sigma_h\sigma}\rangle_{fifth} \approx \delta_{\sigma'_e\sigma_e} \delta_{\sigma'_h\sigma} \delta_{\sigma_h\sigma'} e^{-i(\mathbf{k}-\mathbf{k}')\cdot\mathbf{R}} C^2/A \\
&\quad \sum_{\mathbf{q}'_X \mathbf{k}'_h} W_h^*(\mathbf{k}'_h) \psi_X^*(\mathbf{q}'_X) W_h(\mathbf{k}'_h + \mathbf{k} - \mathbf{k}') \psi_X\left(\mathbf{k}'_h + \frac{\mathbf{k}}{2}\right) V\left(\mathbf{q}'_X + \frac{\mathbf{k}'}{2} - \mathbf{k} - \mathbf{k}'_h\right)
\end{aligned} \tag{8}$$

$$\begin{aligned}
&\langle\Psi_{\mathbf{k}',j\sigma'_e\sigma'_h\sigma'}|H_{eh}|\Psi_{\mathbf{k},j\sigma_e\sigma_h\sigma}\rangle_{seventh} \approx \delta_{\sigma'_e\sigma_e} \delta_{\sigma'_h\sigma} \delta_{\sigma_h\sigma'} e^{-i(\mathbf{k}-\mathbf{k}')\cdot\mathbf{R}} C^2/A \\
&\quad \sum_{\mathbf{q}_X \mathbf{k}_h} W_h^*(\mathbf{k}_h + \mathbf{k}' - \mathbf{k}) \psi_X^*\left(\mathbf{k}_h + \frac{\mathbf{k}'}{2}\right) W_h(\mathbf{k}_h) \psi_X(\mathbf{q}_X) V\left(\mathbf{q}_X + \frac{\mathbf{k}}{2} - \mathbf{k}' - \mathbf{k}_h\right)
\end{aligned} \tag{9}$$

**Exciton mediated spin-spin interaction between trapped holes.** The Hamiltonian for the exciton and hole in the basis of exciton momentum eigenstate and moiré trapped hole is,

$$H_0 = \sum_{\mathbf{k}\sigma_e\sigma_h} \epsilon_{\mathbf{k}\sigma_e\sigma_h} a^\dagger_{\mathbf{k}\sigma_e\sigma_h} a_{\mathbf{k}\sigma_e\sigma_h} + \sum_{j\sigma} \epsilon_d d^\dagger_{j\sigma} d_{j\sigma}$$
$$H_v = \sum_{j\mathbf{k}\mathbf{k}'\sigma\sigma'\sigma_e} \langle \Psi_{\mathbf{k},j\sigma_e\sigma\sigma'} | H_{Coulomb} | \Psi_{\mathbf{k}',j\sigma_e\sigma'\sigma} \rangle d^\dagger_{j\sigma'} d_{j\sigma} a^\dagger_{\mathbf{k}\sigma_e\sigma} a_{\mathbf{k}'\sigma_e\sigma'} \quad (10)$$
$$= \sum_{j\mathbf{k}\mathbf{k}'\sigma\sigma'\sigma_e} I\, e^{-i(\mathbf{k}-\mathbf{k}')\cdot \mathbf{R}_j} d^\dagger_{j\sigma'} d_{j\sigma} a^\dagger_{\mathbf{k}\sigma_e\sigma} a_{\mathbf{k}'\sigma_e\sigma'}$$

where only the exchange part of the exciton hole interaction is included as $H_v$. The exciton mediated spin-spin interaction between trapped holes arises as the second order perturbation of the off-diagonal exciton-hole exchange. And the exciton-hole exchange coefficient $I$ in $H_v$ contains three contributions given in Eq. (7), (8) and (9) respectively. The term from the intralayer hole-hole interaction is given as,

$$I_{hh}A \approx -\frac{64\pi^2}{A^2 a_b^2 a_h^2}\sum_{\mathbf{q}_X \mathbf{k}_h} W_h^*(\mathbf{k}_h - \mathbf{k} + \mathbf{k}') \psi_X^*\left(\mathbf{q}_X + \frac{\mathbf{k}-\mathbf{k}'}{2}\right) W_h(\mathbf{k}_h) \psi_X(\mathbf{q}_X) V\left(\mathbf{q}_X + \frac{\mathbf{k}}{2} - \mathbf{k}' - \mathbf{k}_h\right)$$
$$\approx -\frac{4}{\pi^2 a_b^2 a_h^2} \int d\mathbf{q}_X d\mathbf{k}_h W_h^*(\mathbf{k}_h) \psi_X^*(\mathbf{q}_X) W_h(\mathbf{k}_h) \psi_X(\mathbf{q}_X) \frac{e^2}{4\pi\varepsilon_r\varepsilon_0 (k_{TF} + |\mathbf{q}_X - \mathbf{k}_h|)} \quad (11)$$

In the second line, the dependence on $\mathbf{k}$ and $\mathbf{k}'$ is neglected, because the exciton states involved are the small momentum ones (much smaller compared to the momentum space width of $W_h$ and $\psi_X$). In such limit, both terms from the interlayer electron-hole interaction have the same form,

$$I_{eh}A \approx \frac{4}{\pi^2 a_b^2 a_h^2} \int d\mathbf{q}_X d\mathbf{k}_h W_h^*(\mathbf{k}_h) \psi_X^*(\mathbf{k}_h) W_h(\mathbf{k}_h) \psi_X(\mathbf{q}_X) \frac{e^2}{4\pi\varepsilon_r\varepsilon_0 (k_{TF} + |\mathbf{q}_X - \mathbf{k}_h|)} e^{-d|\mathbf{q}_X - \mathbf{k}_h|} \quad (12)$$

With the selection of the following parameters, $a_b = 2$ nm, $a_h = 2$ nm, $k_{TF} = \frac{10\omega_0}{c} \approx \frac{100}{7}\ \mu\text{m}^{-1}$, $d = 0.7$ nm and $\varepsilon_r = 4$, we obtain $I_{hh}A \approx -0.71$ eV·nm$^2$ and $I_{eh}A \approx 0.5$ eV·nm$^2$. Thus,

$$I = I_{hh} + 2I_{eh} \approx \frac{0.29\ \text{eV·nm}^2}{A} \quad (13)$$

To find the form of the exciton mediated spin-spin interaction between trapped holes, Schrieffer-Wolff transformation is used to eliminate the off-diagonal $H_v$. Introducing,

$$S = -\sum_{j\mathbf{k}\mathbf{k}'\sigma\sigma'\sigma_e} \frac{I e^{-i(\mathbf{k}-\mathbf{k}')\cdot\mathbf{R}_j}}{\epsilon_{\mathbf{k}\sigma_e\sigma} - \epsilon_{\mathbf{k}'\sigma_e\sigma'}} d^\dagger_{j\sigma'} d_{j\sigma} a^\dagger_{\mathbf{k}\sigma_e\sigma} a_{\mathbf{k}'\sigma_e\sigma'} \quad (14)$$

which satisfies $[S, H_0] = -H_v$, the leading perturbation term in the transformed Hamiltonian $e^S H e^{-S}$ is then given as $\Delta H^{(2)} = \frac{1}{2}[S, H_v]$,

$$\Delta H^{(2)} = \sum_{jj'\mathbf{k}\mathbf{k}'\sigma\sigma'\sigma_e} \frac{I^2}{2} \frac{e^{i(\mathbf{k}'-\mathbf{k})\cdot(\mathbf{R}_j - \mathbf{R}_{j'})}}{\epsilon_{\mathbf{k}\sigma_e\sigma} - \epsilon_{\mathbf{k}'\sigma_e\sigma'}} \left[ d^\dagger_{j\sigma'} d_{j\sigma} a^\dagger_{\mathbf{k}\sigma_e\sigma} a_{\mathbf{k}'\sigma_e\sigma'}, d^\dagger_{j'\sigma} d_{j'\sigma'} a^\dagger_{\mathbf{k}'\sigma_e\sigma'} a_{\mathbf{k}\sigma_e\sigma} \right] \quad (15)$$

Tracing out the exciton degrees of freedom with a diagonal density matrix, i.e. $\langle a^\dagger_{\mathbf{k}\sigma_e\sigma} a_{\mathbf{k}\sigma_e\sigma} \rangle = f_{\mathbf{k}\sigma_e\sigma}$,

$$\Delta H^{(2)} = \sum_{j\mathbf{k}\mathbf{k}'\sigma\sigma'\sigma_e} \frac{I^2}{2} \frac{1}{\epsilon_{\mathbf{k}\sigma_e\sigma}-\epsilon_{\mathbf{k}'\sigma_e\sigma'}} \left(f_{\mathbf{k}\sigma_e\sigma} d^\dagger_{j\sigma'} d_{j\sigma'} - f_{\mathbf{k}'\sigma_e\sigma'} d^\dagger_{j\sigma} d_{j\sigma}\right)$$

$$+ \sum_{j\mathbf{k}\mathbf{k}'\sigma\sigma'\sigma_e} \frac{I^2}{2} \frac{1}{\epsilon_{\mathbf{k}\sigma_e\sigma}-\epsilon_{\mathbf{k}'\sigma_e\sigma'}} \left(f_{\mathbf{k}\sigma_e\sigma} f_{\mathbf{k}'\sigma_e\sigma'} d^\dagger_{j\sigma'} d_{j\sigma'} - f_{\mathbf{k}'\sigma_e\sigma'} f_{\mathbf{k}\sigma_e\sigma} d^\dagger_{j\sigma} d_{j\sigma}\right) \quad (16)$$

$$+ \sum_{jj'\mathbf{k}\mathbf{k}'\sigma\sigma'\sigma_e} \frac{I^2}{2} \frac{e^{i(\mathbf{k}'-\mathbf{k})\cdot(\mathbf{R}_j-\mathbf{R}_{j'})}}{\epsilon_{\mathbf{k}\sigma_e\sigma}-\epsilon_{\mathbf{k}'\sigma_e\sigma'}} \left(f_{\mathbf{k}\sigma_e\sigma} - f_{\mathbf{k}'\sigma_e\sigma'}\right) d^\dagger_{j\sigma'} d_{j\sigma} d^\dagger_{j'\sigma} d_{j'\sigma'}$$

where the last term is the mediated spin-spin interaction between the holes (as illustrated in Fig. 3d). We note that the exchange with the trapped hole can flip an exciton between the spin triplet and singlet configurations (of the intravalley and intervalley configurations respectively in the R type WS$_2$/WSe$_2$). So, the denominator involves the energy of the species $\epsilon_{\mathbf{k}\sigma_e\sigma} - \epsilon_{\mathbf{k}'\sigma_e\sigma'}$. As electron-hole exchange is quenched by layer separation, there is no splitting between the intravalley spin triplet and intervalley singlet exciton dispersion, i.e. $\epsilon_\mathbf{k} \equiv \epsilon_{\mathbf{k}\sigma\sigma} = \epsilon_{\mathbf{k}\sigma\bar{\sigma}}$. Thus, the spin-spin interaction terms in $\Delta H^{(2)}$ between the trapped holes is of the form,

$$H_{SS} = \sum_{jj'} J(\mathbf{R}_j - \mathbf{R}_{j'}) \mathbf{S}_j \cdot \mathbf{S}_{j'} \quad (17)$$

where

$$J(\mathbf{R}_j - \mathbf{R}_{j'}) = \sum_{\mathbf{k}\mathbf{k}'} \frac{I^2}{2} \cos[(\mathbf{k}'-\mathbf{k}) \cdot (\mathbf{R}_j - \mathbf{R}_{j'})] \frac{f_\mathbf{k} - f_{\mathbf{k}'}}{\epsilon_\mathbf{k} - \epsilon_{\mathbf{k}'}} \quad (18)$$

**Acknowledgements:** We thank Boris Spivak, Ting Cao, Jiun-Haw Chu, David Cobden, Matt Yankowitz, Cory Dean, and Abhay Narayan Pasupathy for helpful discussions. Research on the observation of ferromagnetism near -1/3 moiré superlattice filling is primarily supported as part of Programmable Quantum Materials, an Energy Frontier Research Center funded by the U.S. Department of Energy (DOE), Office of Science, Basic Energy Sciences (BES), under award DE-SC0019443. Optically induced magnetism of dilute electron/hole gas is mainly supported by DoE BES under award DE-SC0018171. Sample fabrication and PFM characterization are partially supported by ARO MURI program (grant no. W911NF-18-1-0431). The AFM-related measurements were performed using instrumentation supported by the U.S. National Science Foundation through the UW Molecular Engineering Materials Center (MEM·C), a Materials Research Science and Engineering Center (DMR-1719797). WY and CX acknowledge support by the Croucher Foundation (Croucher Senior Research Fellowship) and the University Grant Committee/Research Grants Council of Hong Kong SAR (AoE/P-701/20). Bulk $WSe_2$ crystal growth and characterization by JY is supported by the US Department of Energy, Office of Science, Basic Energy Sciences, Materials Sciences and Engineering Division. KW and TT acknowledge support from the Elemental Strategy Initiative conducted by the MEXT, Japan, Grant Number JPMXP0112101001, JSPS KAKENHI Grant Numbers JP20H00354 and the CREST(JPMJCR15F3), JST. XX acknowledges support from the State of Washington funded Clean Energy Institute and from the Boeing Distinguished Professorship in Physics.

**Author contributions:** XX and WY conceived the project. HP and XW fabricated and characterized the samples, assisted by JZ. XW performed the magneto-optical measurements, assisted by JZ. XW, XX, WY, DX, DRG analyzed and interpreted the results. CX and WY performed calculations of optically induced magnetic exchange interactions. CW and DX performed Monte Carlo simulations of correlated states. TT and KW synthesized the hBN crystals. JY synthesized and characterized the bulk $WSe_2$ crystals. XX, XW, WY, DX, DRG wrote the paper with input from all authors. All authors discussed the results.

**Competing Interests:** The authors declare no competing financial interests.

**Data Availability:** The datasets generated during and/or analyzed during this study are available from the corresponding author upon reasonable request.

Figures:

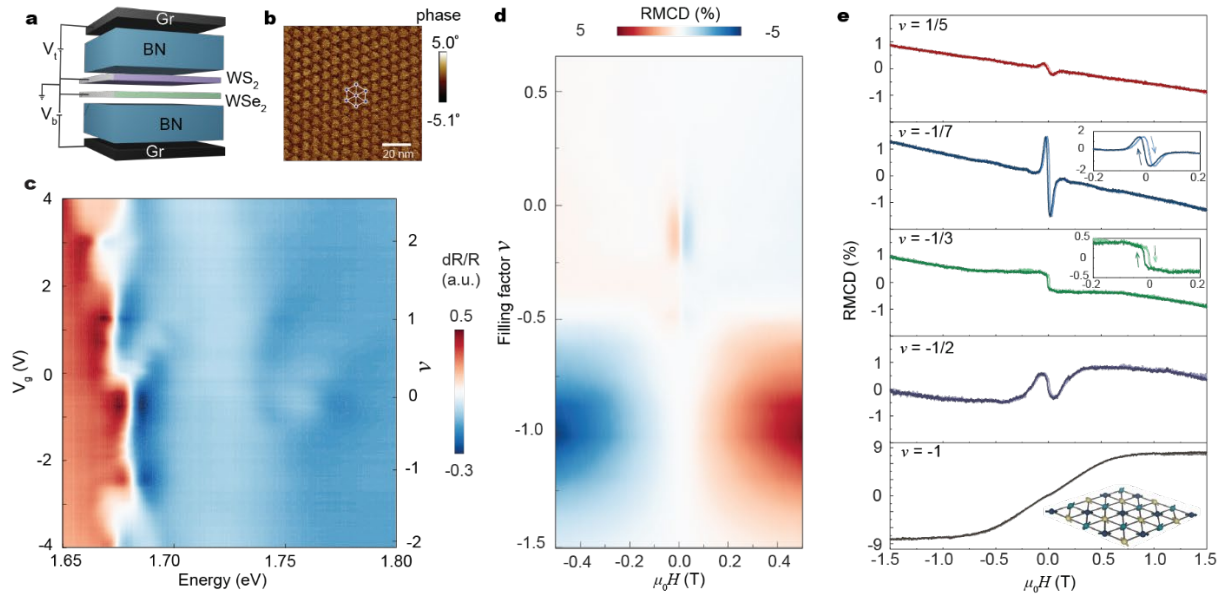

**Figure 1 | Moiré filling dependent magnetic circular dichroism in WS$_2$/WSe$_2$ heterobilayer.**
**a,** Schematic of a dual gated WS$_2$/WSe$_2$ heterobilayer device. **b,** Piezoresponse force microscopy image of a heterobilayer showing triangular moiré superlattice. Scale bar: 20 nm. **c,** Differential optical reflection measurements as a function of doping (gate voltage). Moiré miniband filling factor $v$ is also denoted. **d**, Reflective magnetic circular dichroism (RMCD) signal intensity plotted as a function of filling factor and magnetic fields ($\mu_o H$). **e,** Plots of RMCD signal vs $\mu_o H$ measured sweeping back and forth at selected filling factors. At $v = -1$, a superparamagnetic-like response is observed with 120-degree antiferromagnetic interactions between nearest neighbor hole spins. As doping is reduced, the lineshape of RMCD vs $\mu_o H$ changes dramatically. A sharp feature emerges near zero magnetic field. The associated hysteresis loop, highlighted in the insets, is the hallmark of ferromagnetic order.

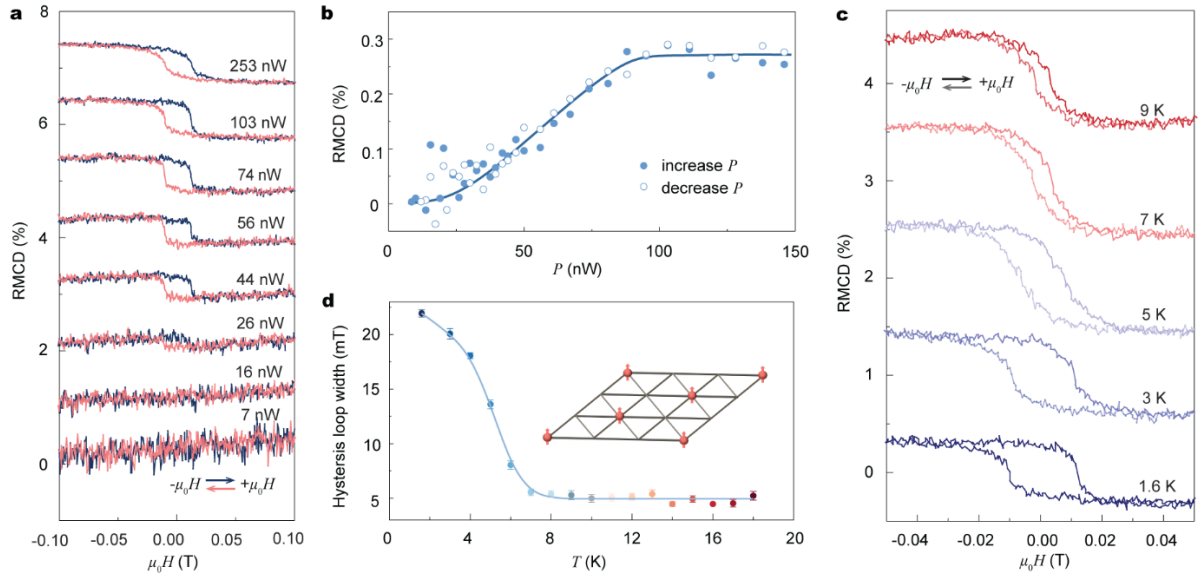

**Figure 2 | Observation of optically induced ferromagnetism near *ν* = -1/3 filling. a,** Power dependent RMCD vs $\mu_o H$ sweeping back and forth at base temperature 1.6 K. The data are offset for clarity. **b**, Remanent RMCD signal at $\mu_o H = 0$ as the optical excitation power increases (solid dots) and decreases (hollow dots). Solid line is a guide to the eye. **c,** Temperature dependent RMCD signal vs $\mu_o H$ at optical excitation power of 103 nW. **d**, Magnetic hysteresis loop width as a function of temperature. Solid line is a guide to the eye. The loop width is determined by the difference between the magnetic fields at which the RMCD signal crosses zero as $\mu_o H$ is swept back and forth. The error bar is the standard deviation obtained by averaging over 5 data points. Inset: Cartoon of ferromagnetic order with spins arranged in a triangular lattice.

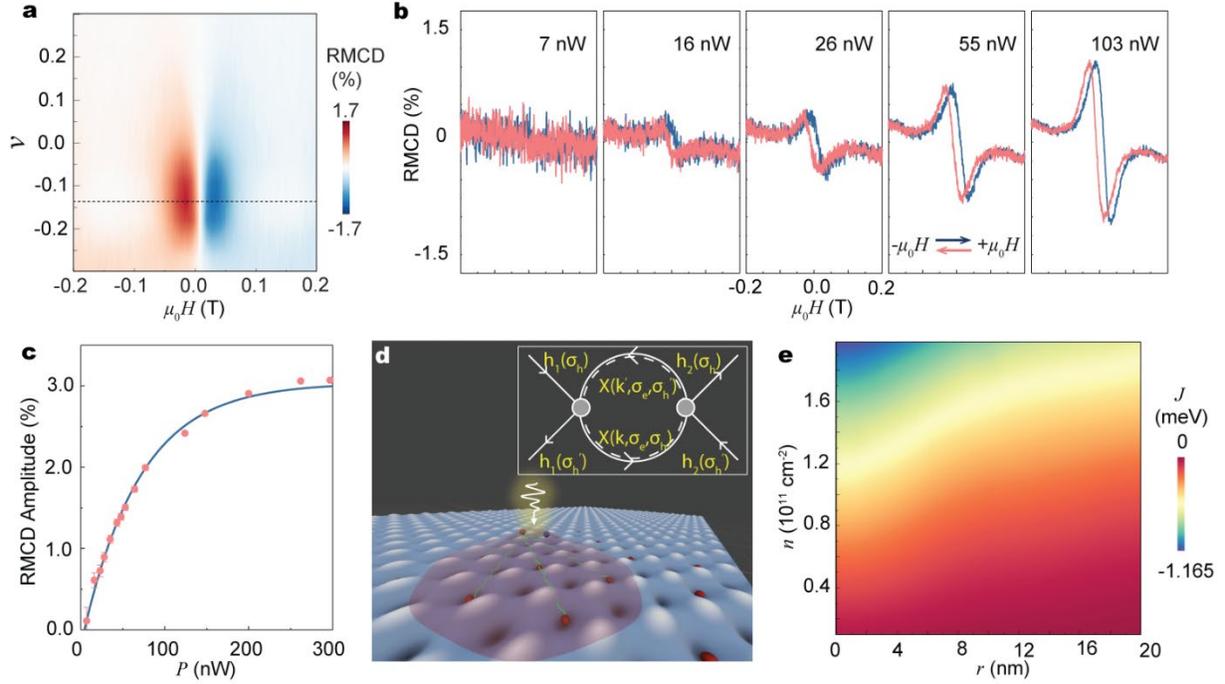

**Figure 3 | Optically induced ferromagnetism in dilute hole gas. a,** RMCD intensity plot at low doping regime. Data is taken at $T$ = 1.6 K and optical excitation power of 200 nW. **b,** Optical excitation power dependent RMCD measurements at fractional filling $v$ = -1/7. **c,** Extracted RMCD amplitude, corresponding to the peak-to-valley value in the heartbeat line shape, versus optical exciton power. Solid line is a guide to the eye. The error bar corresponds to the measurement noise floor of RMCD signal. **d,** Cartoon showing exciton mediated exchanged interactions between moiré trapped spins. The semi-transparent red envelope denotes the center of mass wavefunction of the itinerant exciton. Moiré trapped holes (red dots) interact with each other via a second-order process mediated by their exchange interaction with the hole in the exciton. Inset: a Feynman diagram depicts the spin-spin interactions between two moiré trapped holes mediated by the optically injected exciton. $h_i(\sigma_h)$ denotes a hole at moiré trap $i$ in spin state $\sigma_h$, and $X(k, \sigma_e, \sigma_h)$ denotes an exciton of momentum $k$ with electron/hole spin $\sigma_e/\sigma_h$. **e,** Calculated spin-spin interaction strength $J$ mediated by optically excited exciton as a function of exciton density ($n$) and spatial separation ($r$) between moiré trapped holes. Negative $J$ means ferromagnetic interaction.

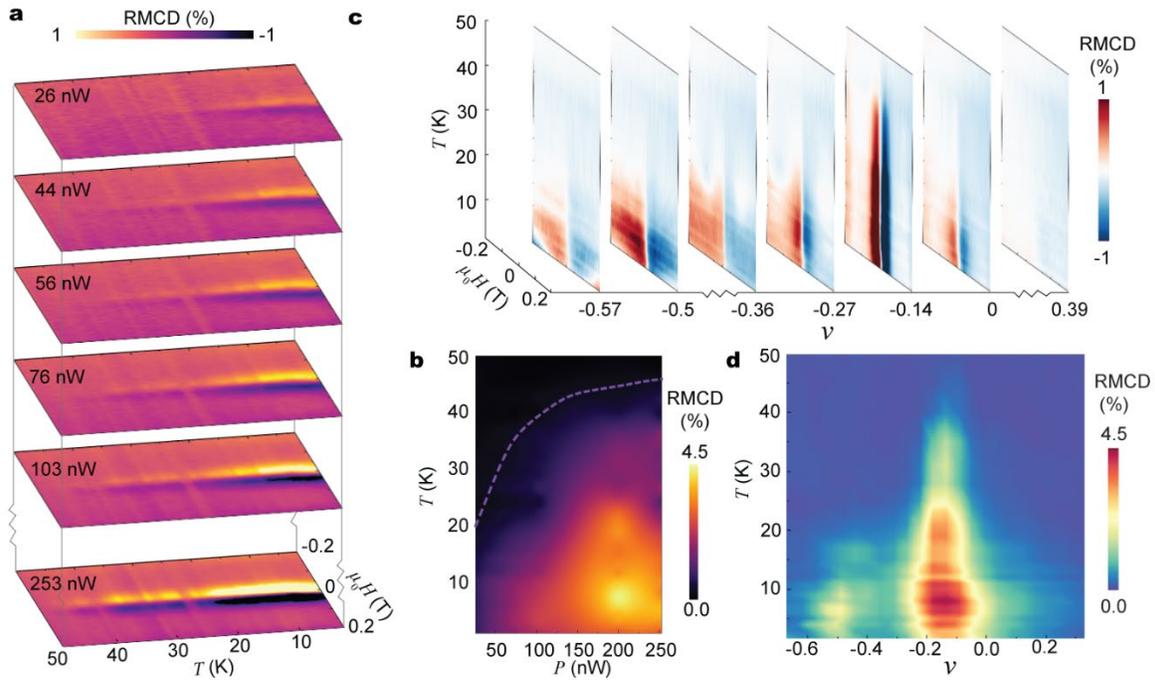

**Figure 4 | Tuning the magnetic states via optical excitation power and filling factor. a,** RMCD signal intensity plots as function of magnetic field and temperature at selected excitation power. The filling factor $v$ is fixed at -1/7. **b**, The extracted RMCD peak-to-valley amplitude as a function of temperature and excitation power at $v$ = -1/7. The critical temperature $T_C$ for the RMCD signal to vanish can be largely tuned by the optical excitation power. Dashed line is a guide to the eye of power dependent $T_C$, which is determined by the 0.2% noise level of RMCD signals. **c,** RMCD signal intensity plots as function of magnetic field and temperature at selected filling factors. **d**, Extracted RMCD signal amplitude as a function of temperature and filling factor, showing the enhanced magnetic response at correlated insulating states at fractional fillings (-1/2, -1/7). The excitation power is fixed at 200 nW for (c) & (d).

# Extended data file
# Light-Induced Ferromagnetism in Moiré Superlattices

Xi Wang[1,2], Chengxin Xiao[3,4], Heonjoon Park[1], Jiayi Zhu[1], Chong Wang[8], Takashi Taniguchi[5], Kenji Watanabe[6], Jiaqiang Yan[7], Di Xiao[1,8], Daniel R. Gamelin[2], Wang Yao[3,4,*], and Xiaodong Xu[1,8,*]

Correspondence to: wangyao@hku.hk; xuxd@uw.edu

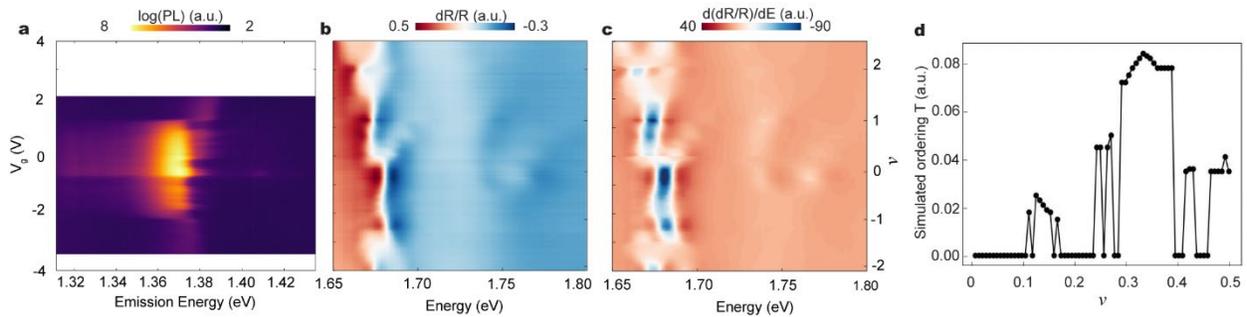

**Extended Data Figure 1. Determination of filling factors. a**, Gate dependent photoluminescence (PL) spectra obtained at 7 K. The excitation wavelength is 740 nm with 50 nW power. Charge neutrality is assigned with the voltage corresponding to the brightest PL with symmetric spectra on the electron and hole sides. **b**, Gate-dependent differential reflectance spectra (ΔR/R) near the WSe$_2$ exciton resonance, measured at base temperature 1.6 K. The power used for reflectance spectroscopy is 15 nW. **c**, Gate-dependent differential reflectance spectra differentiated with respect to photon energy (the same data as in Fig. 1c). All three figures share same y axes. Gate voltages are labeled at the left axis of panel (a). Assigned filling factors are labelled at the right axis of panel (c). The optical doping effect is negligible. The data of **a-c** is taken at a different spot of the same sample. d, Monte Carlo simulation of correlated insulating states at fractional moiré miniband fillings. The transition temperature of charge ordered states in the vertical axis is determined as the temperature where the specific heat is maximum in Monte Carlo simulations.

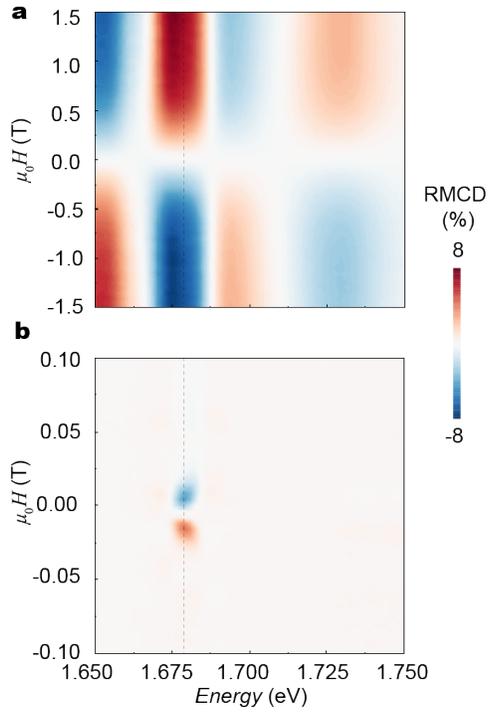

**Extended Data Figure 2. Excitation Energy dependent RMCD at selected filling factors. a,** Excitation Energy dependent RMCD signal at $v = -1$. **b,** Excitation Energy dependent RMCD at $v = -1/7$. Maximized RMCD signals are observed when the excitation energy is between 1.676-1.678 eV (739-740 nm) as indicated by the dash lines.

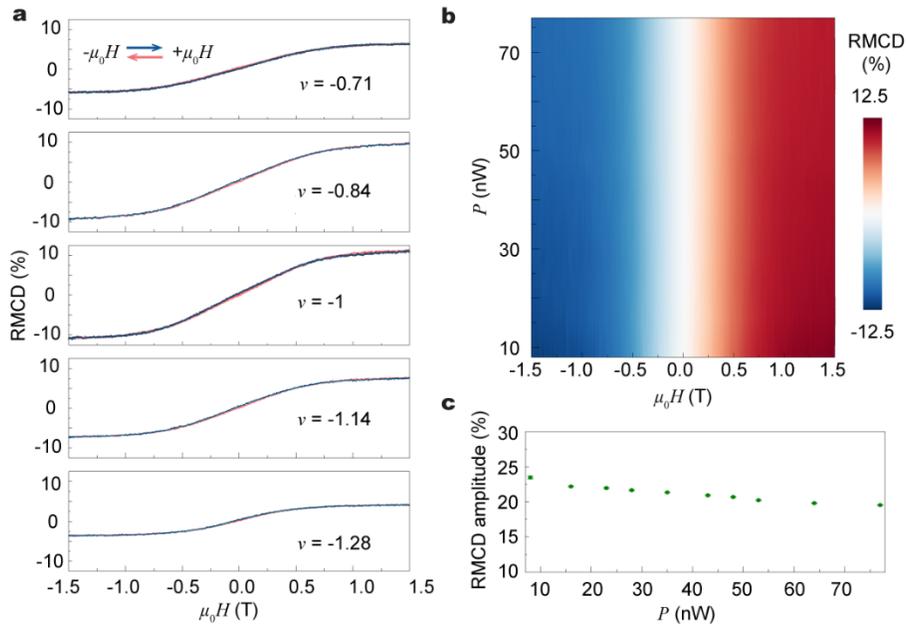

**Extended Data Figure 3. RMCD signal near $v = -1$ and its dependence on optical excitation power. a,** RMCD signal with a few selected filling factors near $v = -1$ at T = 1.6 K. The excitation wavelength is 739.2 nm. Laser power is 24 nW. **b**, Power dependence of RMCD signal at the condition $v = -1$. The RMCD signals barely changes by varying the excitation power. **c**, RMCD amplitude, i.e., saturation RMCD signal difference on the positive and negative sides of magnetic field, as a function of optical excitation power. The saturation amplitude of RMCD signal reduces slightly as optical excitation power increases.

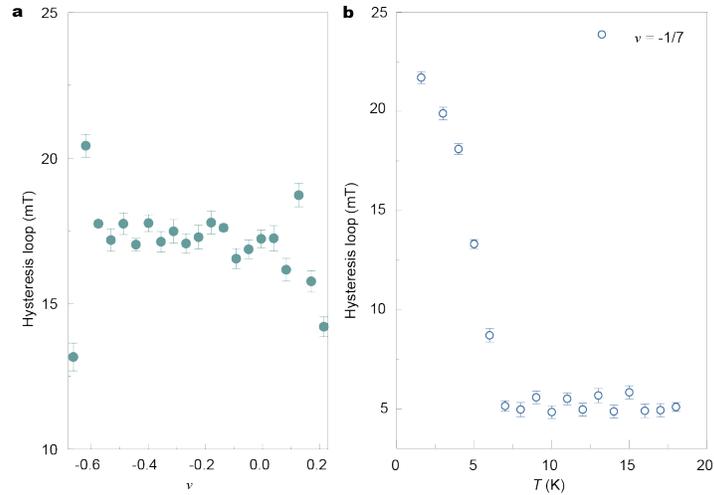

**Extended Data Figure 4. Hysteresis loop width as a function of filling factors and temperature. a,** Extracted hysteresis loop width vs filling factor of the sample in the main text. **b,** Temperature dependent hysteresis loop width of RMCD signal at filling factor $v=-1/7$, showing similar instrument determined offset at high temperature. The loop width is determined by the difference between the magnetic fields at which the RMCD signal crosses zero as $\mu_o H$ is swept back and forth. The error bar is the standard deviation obtained by averaging over 5 data points. The set of data in **a** were taken at different thermal cycles. This causes a slight offset of loop width in (**a**) compared with those at the same filling factors in (**b**) and Fig. 2d in the main text.

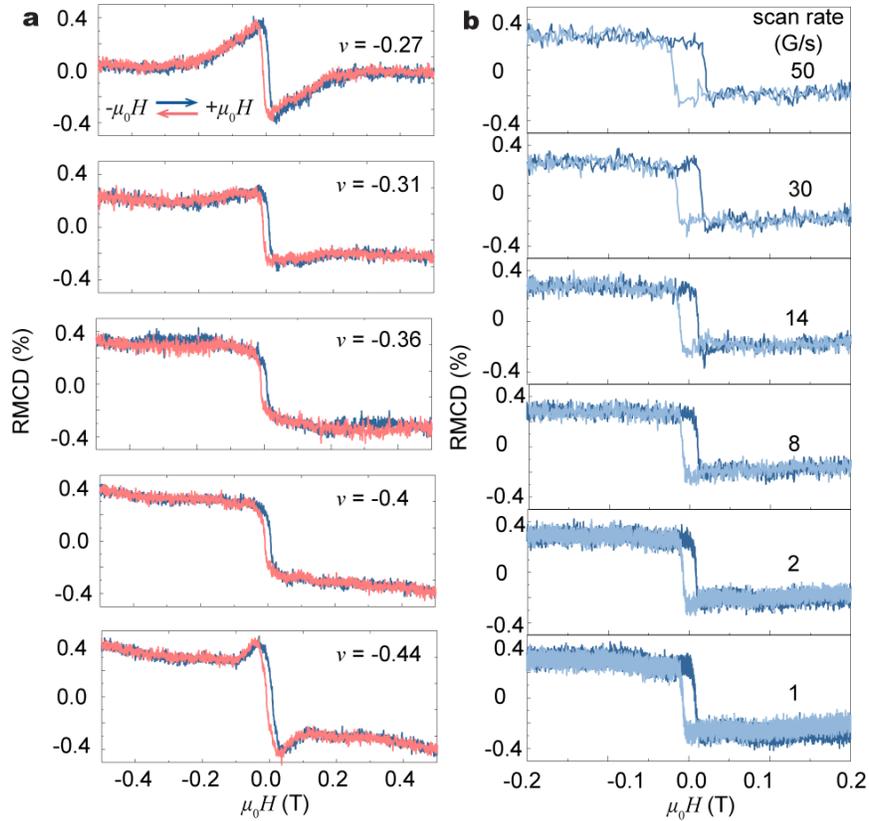

**Extended Data Figure 5. RMCD signal near $v = -1/3$ and its dependence on the magnetic field scanning rate. a.** RMCD signal vs $\mu_o H$ in a narrow doping regime near $v = -1/3$, showing typical ferromagnetic behavior. Data are extracted from Fig. 1d. **b**. Magnetic field sweeping rate dependent RMCD hysteresis loops at $v = -1/3$ with 76 nW excitation.

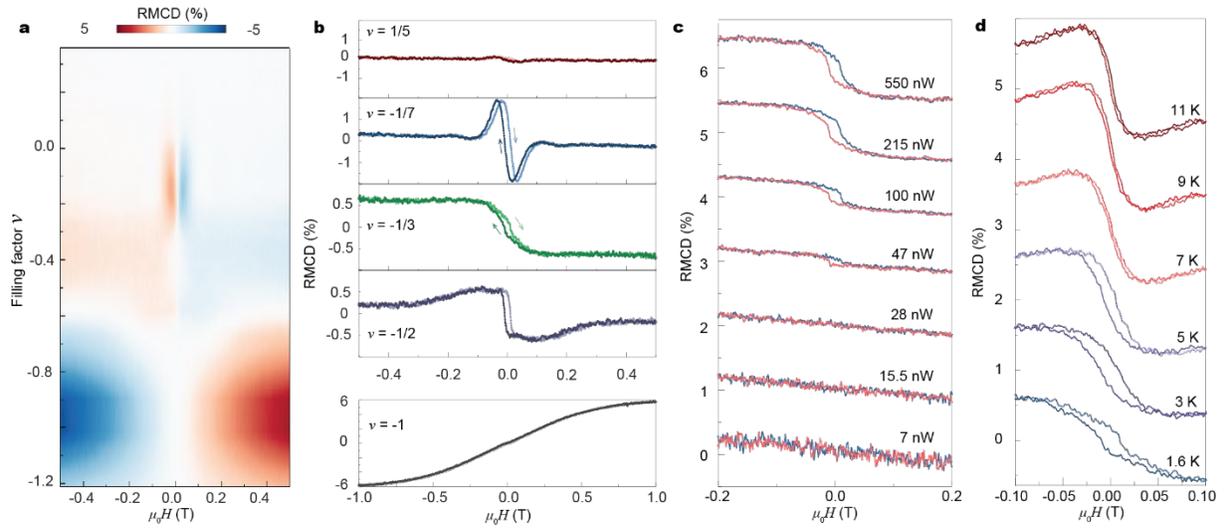

**Extended Data Figure 6. Data from an additional device at near zero twist angle on sapphire substrate.** It resembles the main results observed in the device presented in the maintext. **a,** RMCD signal as a function of filling factor $v$ and magnetic field $\mu_0H$. Temperature: 1.6K. Optical excitation power: 590 nW. **b,** RMCD signal vs $\mu_0H$ measured sweeping back and forth at selected filling factors. **c,** Power dependent RMCD at $v = -1/3$. **d**, Temperature dependent RMCD at $v = -1/3$ and optical excitation power of 590 nW. The data in (**c**) and (**d**) are offset for clarity.

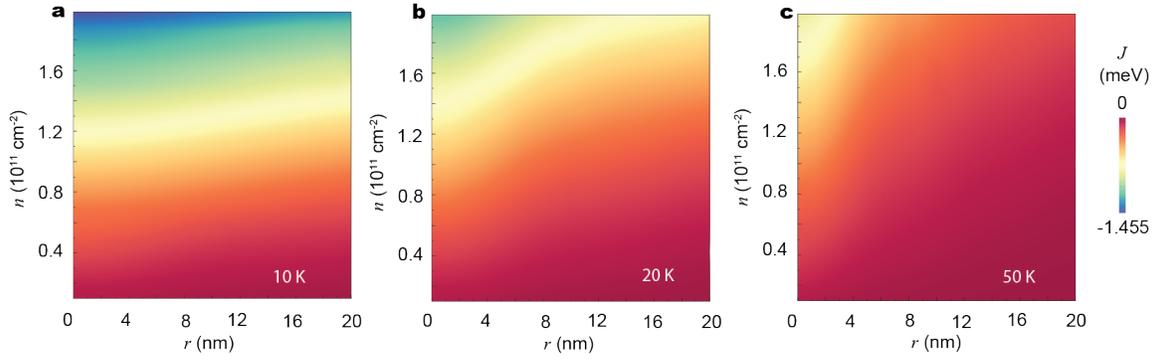

**Extended Data Figure 7. Calculated Exciton mediated exchange interaction *J*. a-c**, shows the *J* amplitude plot as a function of the separation *r* between moiré trapped holes and exciton density with exciton temperature of (a)10 K, (b) 20 K and (c) 50 K.

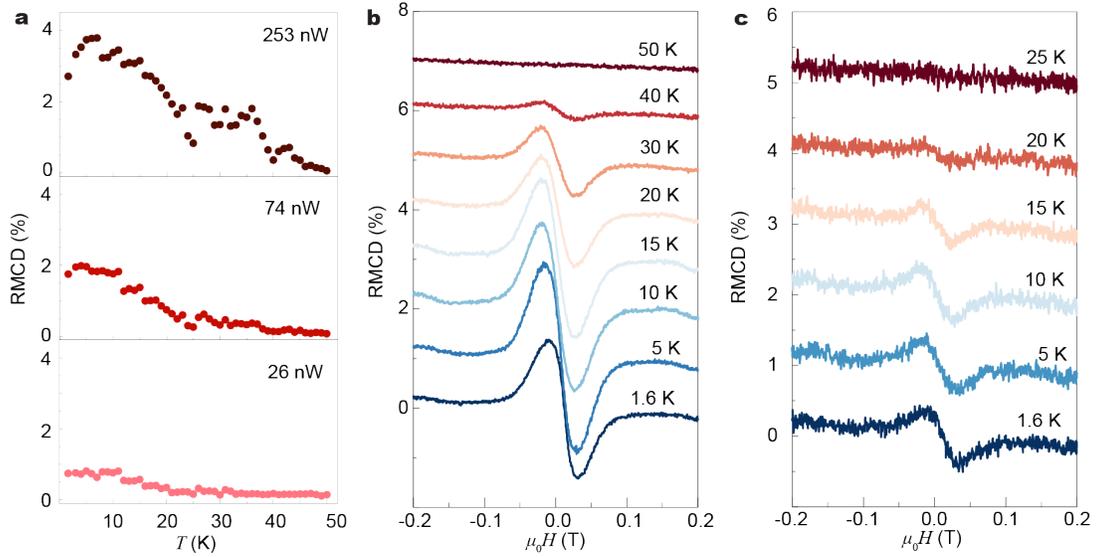

**Extended Data Figure 8. Excitation power dependent RMCD at v=-1/7. a**, Line cuts of Fig. 4b in the main text at three different optical excitation power (indicated in the panels). **b-c,** RMCD signal vs temperature at optical excitation power of (b) 253 nW and (c) 26 nW.

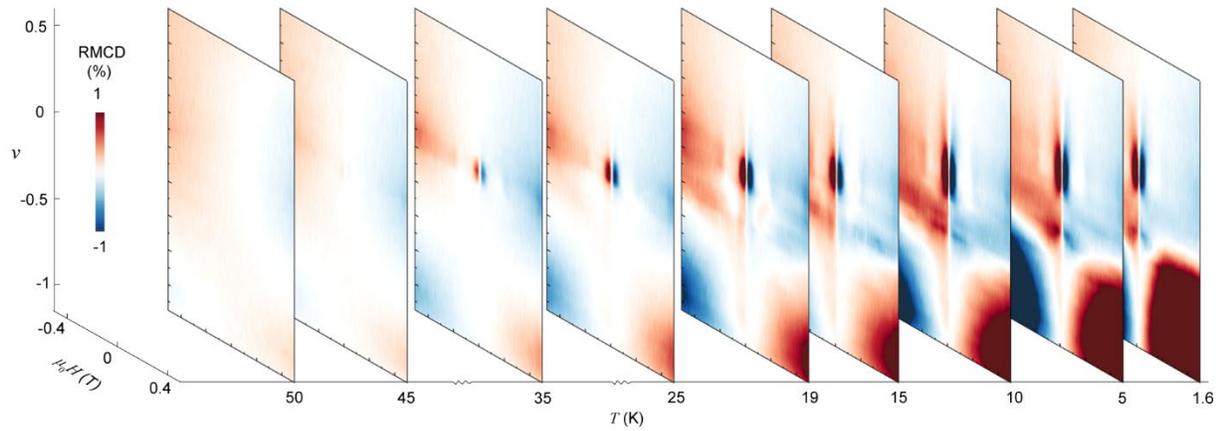

**Extended Data Figure 9. RMCD signal as a function of filling factor and magnetic field at select temperatures.** The excitation power is 200 nW at a wavelength of 739.2 nm.

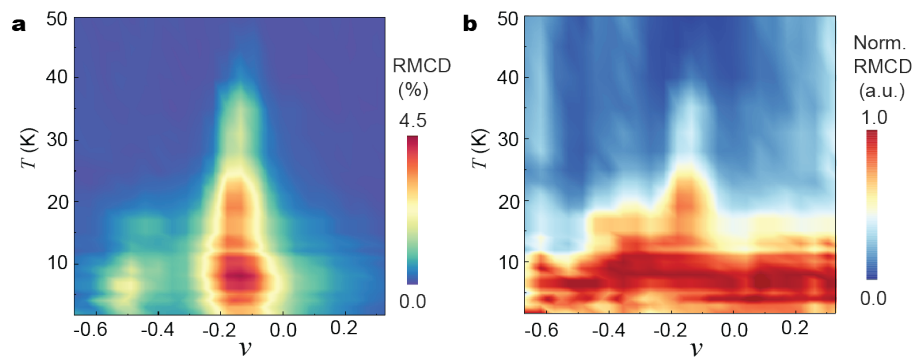

**Extended Data Figure 10. Extracted RMCD signal amplitude as a function of temperature and filling factor. a,** The same plot of Fig. 4d in the main-text. **b,** replot of panel (a) with the RMCD signal normalized to its maximum at each fixed filling factor. The enhanced magnetic response at $v=-1/3$ becomes visible.